\input epsf.tex
\documentclass[12pt]{article}
\usepackage{graphicx}
\usepackage{amssymb,epsf,amsmath}
\csname @addtoreset\endcsname{equation}{section}

\def\bseq{\begin{subequation}}  
\def\eseq{\end{subequation}}
\def\Bar#1{\overline{#1}}                       

\newcommand{\beq}{\begin{equation}}
\newcommand{\eeq}{\end{equation}}
\newcommand{\bea}{\begin{eqnarray}}
\newcommand{\eea}{\end{eqnarray}}
\newcommand{\ena}{\end{eqnarray}}

\newcommand {\non}{\nonumber}

\renewcommand{\a}{\alpha}
\renewcommand{\b}{\beta}

\renewcommand{\d}{\delta}
\renewcommand{\th}{\theta}

\newcommand{\pa}{\partial}
\newcommand{\g}{\gamma}
\newcommand{\G}{\Gamma}

\newcommand{\D}{\Delta}
\newcommand{\e}{\epsilon}

\renewcommand{\l}{\lambda}
\renewcommand{\L}{\Lambda}

\newcommand{\n}{\nu}

\newcommand{\p}{\pi}

\newcommand{\Db}{\Bar{D}}

\newcommand{\thb}{\bar{\theta}}
\newcommand{\Phib}{\bar{\Phi}}

\newcommand{\barh}{\bar{h}}

\newcommand{\Tr}{{\rm Tr}}

\newcommand{\intsup}{\int\!\! d^3xd^4\theta ~}

\newcommand{\intsupk}[1]{\int\!\! \frac{d^3{#1}}{(2\pi)^3} d^4\theta ~}
\newcommand{\intk}[1]{\int\!\! \frac{d^3{#1}}{(2\pi)^3} ~}
\newcommand{\intkk}[2]{\int\!\! \frac{d^3{#1}}{(2\pi)^3} \frac{d^3{#2}}{(2\pi)^3} ~}

\newcommand{\mathN}{\mathcal{N}}
\newcommand{\action}{{\cal S}}

\begin{document}

\begin{titlepage}
{\hbox to\hsize{\hfill December 2009}}

\begin{center}
\vglue 0.99in
{\Large\bf INFRARED STABILITY OF ${\cal N}=2$ \\ [.09in]
CHERN--SIMONS MATTER THEORIES}
\\[.45in]
Marco S. Bianchi \footnote{marco.bianchi@mib.infn.it}, 
Silvia Penati\footnote{silvia.penati@mib.infn.it} ~and~
Massimo Siani\footnote{massimo.siani@mib.infn.it}\\
{\it Dipartimento di Fisica dell'Universit\`a degli studi di
Milano-Bicocca,\\
and INFN, Sezione di Milano-Bicocca, piazza della Scienza 3, I-20126 Milano,
Italy}\\[.8in]

{\bf ABSTRACT}\\[.0015in]
\end{center}
 
According to the AdS4/CFT3 correspondence, ${\cal N}=2$ supersymmetric Chern--Simons matter theories should 
have a stable fixed point in the infrared. In order to support this prediction we  
study RG flows of two--level Chern--Simons matter theories with/without flavors induced by the most 
general marginal superpotential compatible with ${\cal N}=2$ supersymmetry. 
At two loops we determine the complete spectrum of fixed points and study their IR stability. 
Our analysis covers a large class of models including perturbations of the ABJM/ABJ theories with 
and without flavors, ${\cal N}=2,3$ theories with different CS levels corresponding to turning on a 
Romans mass and $\b$--deformations. 
In all cases we find curves (or surfaces) of fixed points which are globally IR stable but locally unstable 
in the following sense: The system has only one direction of stability which in the ABJM case coincides with
the maximal global symmetry preserving perturbation, whereas along any other direction it flows to a different 
fixed point on the surface. 
The question of conformal invariance vs. finiteness is also addressed: While in general vanishing beta--functions
imply two--loop finiteness, we find a particular set of flavored theories where this is no longer true.

${~~~}$ \newline

\end{titlepage}

\section{Introduction}

Recently, a renew interest in three dimensional Chern--Simons (CS) theories has been triggered by the
formulation of the ${\rm AdS}_4/{\rm CFT}_3$ correspondence between CS matter theories and M/string theory. 
While pure CS is a topological theory \cite{Schwarz,Witten}, the addition of 
matter degrees of freedom makes it dynamical and can be used to describe nontrivial 3D systems. The addition 
of matter can be also exploited to formulate theories with extended supersymmetry \cite{schwarz,GY,GW}. Chern--Simons matter 
theories corresponding to a single gauge group can be at most ${\cal N}=3$ supersymmetric \cite{KaoLee}, 
while the use of direct products of groups and matter in the bifundamental representation allows to increase  
supersymmetry up to ${\cal N}=8$ \cite{BLG}. 

This has led to the precise formulation of the ${\rm AdS}_4/{\rm CFT}_3$ correspondence which in its original form \cite{ABJM} states
that M-theory on ${\rm AdS}_4 \times S^7/{\cal Z}_k$ describes the strongly coupled dynamics of a two--level ${\cal N}=6$ 
supersymmetric Chern--Simons theory with $U(N)_k \times U(N)_{-k}$ gauge group and $SU(2) \times SU(2)$
invariant matter in the bifundamental.  This is the field theory generated at low energies by a stack of $N$ M2--branes
probing a ${\cal C}^4/{\cal Z}_k$ singularity.
In the decoupling limit $N \to \infty$ with $\l \equiv N/k$ large and fixed, choosing $N \ll k^5$, the radius of the eleventh dimension
in M-theory shrinks to zero and the dual description is given in terms of a type IIA string theory on 
${\rm AdS}_4  \times {\cal CP}^3$ background \cite{ABJM}.    
In the particular case of $N=2$, supersymmetry gets enhanced to ${\cal N}=8$ and the strings provide 
a dual description of the Bagger--Lambert--Gustavsson (BLG) model \cite{BLG,gustavsson,VR}. Enhancement of supersymmetry 
occurs also for $k= 1,2$ where the ABJM theory describes the low energy dynamics of $N$ membranes in flat space and
in  ${\cal R}^8/{\cal Z}_2$, respectively \cite{ABJM,GR, KOS}.
 
Since the original formulation of the correspondence, a lot of work has been done for studying 
the dynamical properties of this particular class of CS matter theories, such as integrability 
\cite{NT}-\cite{BMR}, the structure of the chiral ring and the operatorial content \cite{ABJM}, 
\cite{BT}-\cite{KM} 
and dynamical supersymmetry breaking \cite{GKL,AS}. Many efforts have been also devoted to the 
generalization of the correspondence to different gauge groups \cite{ABJ,ST}, 
to less (super)symmetric backgrounds \cite{Klebanov}-\cite{BILPSZ} and to include flavor degrees of freedom 
\cite{HK, GJ, HLT, BCC,  Jafferis}. Theories with two different CS levels $(k_1, k_2)$ have been also introduced 
\cite{schwarz, GT} which correspond to turning on a Romans mass \cite{romans} in the dual background. 

CS matter theories involved in the ${\rm AdS}_4/{\rm CFT}_3$ correspondence are of course at their superconformal fixed 
point \footnote{A classification of a huge landscape of superconformal Chern--Simons matter theories in terms
of matter representations of global symmetries has been given in \cite{MFM}.}. Compactification of type IIA  
supergravity on ${\rm AdS}_4 \times {\cal CP}^3$ does not contain scalar tachyons \cite{NP}. Since these states are dual
to relevant operators in the corresponding field theory, ${\rm AdS}_4/{\rm CFT}_3$ correspondence  leads to the prediction that 
in the far IR fixed points should be stable. 

As a nontrivial check of the correspondence, it is then interesting to investigate the properties of these fixed points 
in the quantum field theory in order to establish whether they are isolated fixed points or they belong to a continuum 
surface of fixed points,
whether they are IR stable and which are the RG trajectories which intersect them. Since for $k \gg N$
the CS theory is weakly coupled, a perturbative approach is available.      

With these motivations in mind, we consider a ${\cal N}=2$ supersymmetric two--level CS theory 
for gauge group $U(N) \times U(M)$ with matter in the bifundamental representation and flavor 
degrees of freedom in the fundamental, perturbed by the most general matter superpotential compatible 
with ${\cal N}=2$ supersymmetry.
For particular values of the couplings the model reduces to the ${\cal N}=6$ ABJ/ABJM superconformal theories 
\cite{ABJM, ABJ} (${\cal N}=8$ BLG theory \cite{BLG} for $N = M = 2$) or to the superconformal ${\cal N}=2,3$ 
theories with different CS levels studied in \cite{GT}, in all cases with and without flavors.
More generally, it describes marginal (but not exactly marginal) perturbations which can drive the theory away from the superconformal points. 
 
At two loops, we compute the beta--functions and determine the spectrum of fixed points. 
In the absence of flavors the condition of vanishing beta--functions necessarily implies the vanishing 
of anomalous dimensions for all the elementary fields of the theory. Therefore, the set of superconformal 
fixed points coincides with the set of superconformal finite theories.
When flavors are present this is no longer true and in the space of the couplings we determine a surface 
of fixed points where the theory is superconformal but not two--loop finite. 

When flavors are turned off we determine a continuum surface of fixed points which contains 
as non--isolated fixed points the BLG, the ABJ and ABJM theories. 
The case of theories with equal CS levels and $U(1)_A \times U(1)_B$ symmetry preserving perturbations has been already investigated in \cite{us}. The present paper provides details for that class of theories and generalizes the results to the case of
no--symmetry preserving perturbations. 
When the CS levels are different the surface contains a ${\cal N}=2$, $SU(2)_A \times SU(2)_B$ invariant and a ${\cal N}=3$ superconformal theories. This result confirms the existence of the superconformal points conjectured in \cite{GT}.
Moreover, we prove that the two theories are connected by a line of ${\cal N}=2$ fixed points, as conjectured there.
 
We extend our analysis to the case of complex couplings, so including fixed points corresponding to beta--deformed theories
\cite{Imeroni}. 

In the presence of flavor matter the spectrum of fixed points spans a seven dimensional hypersurface 
in the space of the couplings which contains the fixed point corresponding to the ABJM/ABJ models 
with flavors studied in \cite{HK, GJ, HLT}. More generally, we find a fixed point which describes a 
${\cal N}=3$ theory with different CS levels with the addition of flavor degrees of freedom \cite{GJ}. 
As a generalization of the pattern arising in the unflavored case, we find that it is connected by a  
four dimensional hypersurface of ${\cal N}=2$ fixed points to a line of ${\cal N}=2$ fixed points 
with $SU(2)_A \times SU(2)_B$ invariance in the bifundamental sector. 
   
We then study RG trajectories around these fixed points in order to investigate their IR stability.
The pattern which arises is common to all these theories, flavors included or not, and can be summarized 
as follows. 

\begin{itemize}
\item
Infrared stable fixed points always exist and we determine the RG trajectories which connect them
to the UV stable fixed point (free theory). 

\item
In general these fixed points belong to a continuum surface. The surface is globally stable 
since RG flows always point towards it. 

\item
Locally, each single fixed point has only one direction of stability which corresponds to perturbations
along the RG trajectory which intersects the surface at that point. 
In the ABJ/ABJM case this direction coincides with the maximal flavor symmetry preserving perturbation
\cite{us}.
Along any other direction, perturbations drive the system away from the original point towards a different point on the 
surface. This is what we call local instability. 
 
\item
When flavors are added, stability is guaranteed by the presence of nontrivial interactions between flavors
and bifundamental matter. The fixed point corresponding to setting these couplings to zero is in fact unstable. 
\end{itemize}

The organization of the paper is the following: In Section 2 we introduce our general theory, we discuss
its properties and quantize it in a manifest ${\cal N}=2$ set-up. In Section 3 we compute the two--loop divergences, 
we renormalize the theory
and determine the beta--functions. Sections 4 and 5 contain the main results of the paper concerning the 
determination of the spectrum of fixed points and the study of the IR stability for the most 
interesting cases. A concluding discussion follows. In the Appendix we list our conventions for  
3D ${\cal N}=2$ superspace.

\section{${\cal N}=2$ Chern--Simons matter theories}

In three dimensions, we consider a ${\cal N}=2$ supersymmetric $U(N) \times U(M)$ Chern--Simons theory  
for vector multiplets $(V,\hat{V})$ coupled to  chiral multiplets $A^i$ and $B_i$, $i=1,2$,  in the $(N,\bar{M})$
and $(\bar{N},M)$ representations of the gauge group respectively, and flavor matter described by two couples of  
chiral superfields $Q_i, \tilde Q_i, \, i=1,2$ charged under the gauge groups and under a global
$U(N_f)_1 \times U(N_f^\prime)_2$. 

The vector multiplets $V, \hat{V}$ are in the adjoint representation of the gauge groups $U(N)$ and $U(M)$
respectively, and we write $V_{\;\; a}^b \equiv V^A (T_A)_{\, \, a}^b$ and $\hat{V}_{\; \; \hat{a}}^{\hat{b}} \equiv 
\hat{V}^A (T_A)_{\, \, \hat{a}}^{\hat{b}}$.  Bifundamental matter carries global $SU(2)_A \times SU(2)_B$ 
indices $A^i, \bar A_i, B_i, \bar B^i$ and local $U(N) \times U(M)$ indices $A^a_{\ \hat a}, 
\bar A^{\hat a}_{\ a}, B^{\hat a}_{\ a}, \bar  B^a_{\ \hat a}$. Flavor matter carries (anti)fundamental 
gauge and global indices, $(Q_1^r)^a, (\tilde{Q}_{1,r})_a, (Q_2^{r'})^{\hat{a}}, (\tilde{Q}_{2,r'})_{\hat{a}}$, 
with $r=1, \cdots N_f$, $r'=1, \cdots N'_f$.

In ${\cal N}=2$ superspace the action reads (for superspace conventions see Appendix)
\beq
{\cal S} = {\cal S}_{\mathrm{CS}} + {\cal S}_{\mathrm{mat}} +
  {\cal S}_{\mathrm{pot}} 
  \label{eqn:action} 
\eeq
with
\bea  
{\cal S}_{\mathrm{CS}}
  &=& K_1 \int d^3x\,d^4\theta \int_0^1 dt\: \Tr \Big[
  V \Db^\a \left( e^{-t V} D_\a e^{t V} \right) \Big] 
  \non \\
  &+& 
  K_2 \int d^3x\,d^4\theta \int_0^1 dt\: \Tr \Big[
  \hat{V} \Db^\a \left( e^{-t \hat{V}} D_\a
  e^{t \hat{V}} \right) \Big]   
  \label{eqn:CS-action} 
  \\ 
  \non \\
  {\cal S}_{\mathrm{mat}} &=& \int d^3x\,d^4\theta\: \Tr \left( \bar{A}_i
  e^V A^i e^{- \hat{V}} + \bar{B}^i e^{\hat V} B_i
  e^{-V} \right) 
  \non \\
  &+& \intsup \Tr \left( {\bar Q}^1_r e^V Q^r_1 +
  \bar{\tilde{Q}}^{1,r} \tilde Q_{1,r} e^{-V} + \bar Q^2_{r^\prime}
  e^{\hat V} Q_2^{r^\prime} + \bar{\tilde Q}^{2, r^\prime} \tilde Q_{2,
  r^\prime} e^{-\hat V} \right)
    \label{eqn:mat-action} \\ 
  \non \\  
  {\cal S}_{\mathrm{pot}} &=& \int d^3x\,d^2\theta\:
   \Tr  \left[ h_1 (A^1 B_1)^2 + h_2 (A^2 B_2)^2 + h_3 (A^1 B_1 A^2 B_2)
  + h_4 (A^2 B_1 A^1 B_2) \right.
  \non \\ 
  && \left. + \l_1 (Q_1 \tilde Q_1)^2 + \l_2
  (Q_2 \tilde Q_2)^2 + \l_3 Q_1 \tilde Q_1 Q_2 \tilde Q_2 \right.
   \label{eqn:superpot}\\ 
  &&
  \left. + \a_1 \tilde Q_1 A^1 B_1 Q_1 + \a_2 \tilde Q_1 A^2 B_2 Q_1
  + \a_3 \tilde Q_2 B_1 A^1 Q_2 + \a_4 \tilde Q_2 B_2 A^2 Q_2 \right]  \, + \, h.c.
  \non
\eea
Here $2\pi K_1$, $2\pi K_2$ are two independent integers, as required by gauge invariance of the effective action. 
In the perturbative regime we take $K_1, K_2 \gg N,M$. The superpotential 
(\ref{eqn:superpot}) is the most general classically marginal perturbation which 
respects ${\cal N}=2$ supersymmetry but allows only for a $U(N_f) \times U(N^\prime_f)$ global symmetry 
in addition to a global $U(1)$ under which the bifundamentals have for example charges $(1,0,-1,0)$. 

For generic values of the couplings, the action ({\ref{eqn:action}) is invariant under the 
following gauge transformations
\bea
&& e^V \rightarrow e^{i \bar{\L}_1} e^V e^{-i\L_1} 
\qquad \qquad e^{\hat{V}} \rightarrow e^{i\bar{\L}_2} e^{\hat{V}} e^{-i\L_2} 
\\
&& \non \\
&& A^i \rightarrow e^{i\L_1} A^i e^{-i\L_2}
\qquad \qquad B_i \rightarrow e^{i\L_2} B_i e^{-i\L_1}
\non \\
&& Q_1 \rightarrow e^{i\L_1} Q_1 
\qquad \qquad \qquad \tilde{Q}_1 \rightarrow  \tilde{Q}_1 e^{-i\L_1}
\non \\
&& Q_2 \rightarrow e^{i\L_2} Q_2 
\qquad \qquad \qquad \tilde{Q}_2 \rightarrow  \tilde{Q}_2 e^{-i\L_2} 
\label{gaugetransf}
\eea
where $\L_1, \L_2$ are two chiral superfields parametrizing $U(N)$ and $U(M)$ gauge transformations,
respectively. Antichiral superfields transform according to the conjugate of (\ref{gaugetransf}).

For special values of the couplings we can have enhancement of global symmetries
and/or R--symmetry with consequent enhancement of supersymmetry. We list the most important
cases we will be interested in. 

\vskip 10pt
\noindent
\underline{Theories without flavors} 

\vskip 10pt
\noindent 
Turning off flavor matter ($N_f = N'_f = 0$, $\a_j = \l_j = 0$) and setting 
\beq
K_1 = - K_2 \equiv K  \qquad , \qquad h_1 = h_2 = 0  
\eeq
we have ${\cal N}=2$ ABJM/ABJ--like theories already studied in \cite{us}. In this case the theory is invariant
under two global $U(1)$'s in addition to $U(1)_R$. The transformations are
\bea
&& U(1)_A: \quad A^1 \rightarrow e^{i\a} A^1  \qquad ~, \qquad  U(1)_B: \quad B_1 \rightarrow e^{i\b} B_1
\non \\
&& ~ \qquad \qquad A^2 \rightarrow e^{-i\a} A^2 \qquad , \qquad \qquad  \qquad ~ B_2 \rightarrow e^{-i\b} B_2
\label{U(1)}
\eea
When $h_3 = - h_4 \equiv h$, the global symmetry becomes $U(1)_R \times SU(2)_A 
\times SU(2)_B$ and gets enhanced to $SU(4)_R$ for $h = 1/K$ \cite{ABJM, Klebanov}. For this particular
values of the couplings we recover the ${\cal N}=6$ superconformal ABJ theory \cite{ABJ} and for 
$N=M$ the ABJM theory \cite{ABJM}.

More generally, we can select theories corresponding to complex couplings
\beq
\qquad h_3 = h e^{i\pi \b} \quad , \quad  h_4 = -h e^{-i\pi \b} 
\eeq
These are ${\cal N}=2$ $\b$--deformations of the ABJ--like theories. For particular values of $h$ and
$\b$ we find a superconformal invariant theory.

\vskip 12pt
\noindent
Going back to real couplings, we now consider the more general case $K_1 \neq -K_2$. Setting 
\beq
h_1 = h_2 = \frac12 \left( h_3 + h_4 \right) 
\eeq
the corresponding superpotential reads
\beq
{\cal S}_{\mathrm{pot}} = \frac12 \, \int d^3x\,d^2\theta\:
   \Tr  \left[ h_3 (A^i B_i)^2 + h_4 (B_i A^i)^2 \right] \, + \, h.c.
\label{potGT}
\eeq
This is the class of ${\cal N}=2$ theories studied in \cite{GT} with $SU(2)$ invariant superpotential,
where $SU(2)$ rotates simultaneously $A^i$ and $B_i$.

When $h_3 = -h_4$, that is $h_1 = h_2 = 0$,
we have the particular set of ${\cal N}=2$ theories with global $SU(2)_A \times SU(2)_B$ symmetry
\cite{GT}. This is the generalization of ABJ/ABJM--like theories to $K_1 \neq -K_2$.
According to AdS/CFT, for particular values of $h_3= -h_4$ we should find a superconformal invariant theory. 

Another interesting fixed point should correspond to $h_3 = \frac{1}{K_1}$ and $h_4 = \frac{1}{K_2}$.
The $U(1)_R$ R--symmetry is enhanced to $SU(2)_R$ and the theory is ${\cal N}=3$ superconformal \cite{GT}.

\vskip 15pt
\noindent
\underline{Theories with flavors}

\vskip 10pt
\noindent
Setting 
\bea
&& K_1 = - K_2 \equiv K \qquad, \qquad  h_1 = h_2 = 0 \qquad , \qquad h_3 = - h_4 = \frac{1}{K} 
\non \\
&& \l_1 = \frac{a_1^2}{2K} \qquad, \qquad \l_2 = -\frac{a_2^2}{2K} \qquad , \qquad \l_3 = 0
\non \\
&& \a_1 = \a_2 = \frac{a_1}{K} \qquad , \qquad \a_3 = \a_4 =  \frac{a_2}{K}   
\eea
with $a_1, a_2$ arbitrary, our model reduces to the class of ${\cal N}=2$ theories studied in 
\cite{HK}. 
Choosing in particular $a_1 = -a_2 = 1$ there is an enhancement of R--symmetry and the theory 
exhibits ${\cal N}=3$ supersymmetry. This set of couplings should correspond to a superconformal 
fixed point \cite{HK, GJ, HLT}. 

In the more general case of $K_1 \neq -K_2$, in analogy with the unflavored case we consider the class 
of theories with
\beq
h_1 = h_2 = \frac12 \left( h_3 + h_4 \right) 
\qquad ; \qquad
\a_1 = \a_2 \quad ,  \quad \a_3 = \a_4 
\label{potGTflavor}
\eeq
For generic couplings these are ${\cal N}=2$ theories with a $SU(2)$ symmetry in the bifundamental sector
which rotates simultaneously $A^i$ and $B_i$. When $h_3 = - h_4$ this symmetry is enhanced to 
$SU(2)_A \times SU(2)_B$. The flavor sector has only $U(N_f) \times U(N_{f}^\prime)$ flavor symmetry.

Within this class of theories we can select the one corresponding to 
\bea
&&\l_1 = \frac{h_3}{2} \qquad ,\qquad \l_2 = \frac{h_4}{2} \qquad ,\qquad \l_3 = 0 
\non \\
&&\quad \a_1 = \a_2 = h_3  \qquad ,\qquad \a_3 = \a_4= h_4
\eea
The values $h_3 = \frac{1}{K_1}, h_4 = \frac{1}{K_2}$ give the ${\cal N}=3$ superconformal theory with 
flavors mentioned in \cite{GJ}. It corresponds to flavoring the ${\cal N}=3$ theory of \cite{GT}.

\vskip 20pt
We now proceed to the quantization of the theory in a manifest ${\cal N}=2$ setup. 

\noindent 
In each gauge sector we choose gauge-fixing functions $\bar{F} = D^2 V$, $F = \bar{D}^2 V$ and 
insert into the functional integral the factor
\beq
\int {\cal D}f {\cal D}\bar{f} \, \D(V) \D^{-1} (V) \, \exp\Big\{ - \frac{K}{2\a} \int d^3x d^2\th \Tr 
(ff) - \frac{K}{2\a} \int d^3x d^2\thb \Tr (\bar{f}\bar{f}) \Big\}
\eeq
where $\D(V) = \int d\L d\bar{\L} \d(F(V, \L, \bar{\L}) - f) \d(\bar{F}(V, \L, \bar{\L}) - \bar{f})$
and the weighting function has been chosen in order to have a dimensionless gauge parameter $\a$.
We note that the choice of the weighting function is slightly different from the four dimensional
case \cite{superspace} where we usually use $\int {\cal D}f {\cal D}\bar{f} \exp\left\{ - \frac{1}{g^2\a} \int d^4x d^4\th 
\Tr (f\bar{f}) \right\}$.
 
The quadratic part of the gauge--fixed action reads
\bea
S_{CS} + S_{gf} &\rightarrow& \frac12 K_1 \intsup \Tr \, V \left( \Db^\a D_\a + \frac{1}{\a}
  D^2 + \frac{1}{\a} \Db^2 \right) V        
\non \\
&~~& + \frac12 K_2 \intsup \Tr \, \hat V \left( \Db^\a D_\a
  + \frac{1}{\a} D^2 + \frac{1}{\a} \Db^2 \right) \hat V 
\eea
and leads to the gauge propagators
\bea
  \langle V^A(1) \, V^B(2) \rangle
   = -\frac{1}{K_1} \frac{1}{\Box} \left( \Db^\a D_\a + \a D^2
   + \a \Db^2 \right) \, \delta^4(\th_1-\th_2) \, \delta^{AB} \\ \langle \hat
   V^A(1) \, \hat V^B(2) \rangle =
   -\frac{1}{K_2} \frac{1}{\Box} \left( \Db^\a D_\a + \a D^2
   + \a \Db^2 \right) \, \delta^4(\th_1-\th_2) \, \delta^{AB}
\label{gaugeprop}
\eea
In our calculations we will use the analog of the Landau gauge, $\a = 0$.

Expanding $S_{CS} + S_{gf}$ at higher orders in $V, \hat{V}$ we obtain the interaction vertices.
For two--loop calculations we need
\bea
S_{CS} + S_{gf} &\rightarrow&  \frac{i}{6} K_1
  f^{ABC} \intsup \left( \Db^\a V^A\, V^B\, D_\a V^C \right) 
  \non \\
  &~& - \frac{1}{24} K_1 f^{ABE} f^{ECD} \intsup \left( \Db^\a V^A\,
  V^B\, D_\a V^C\, V^D \right) 
  \non \\  
  &~& +\frac{i}{6} K_2 f^{ABC} \intsup \left( \Db^\a \hat V^A\, \hat V^B\,
  D_\a \hat V^C \right) 
  \non \\ 
  &~& -\frac{1}{24} K_2 f^{ABE}
  f^{ECD} \intsup \left( \Db^\a \hat V^A\, \hat V^B\, D_\a \hat
  V^C\, \hat V^D \right)
\eea
The ghost action is the same as the one of the four dimensional ${\cal N}=1$ case \cite{superspace}
\bea \label{ghostexp}
  S_{gh} &=&  \Tr \int d^4 x d^4 \theta \left[ \overline{c}' c - c'
  \overline{c} + \frac{1}{2} (c' + \overline{c}') [ V, (c +
  \overline{c})] \right] + {\cal O}(V^2)
\eea
and gives ghost propagators
\beq
  \langle \overline{c}'(1) \, c(2) \rangle = \langle c'(1) \, \overline{c}(2) \rangle =
  - \frac{1}{\Box} \delta^4(\th_1 - \th_2) 
\eeq
and cubic interaction vertices
\beq
  \frac{i}{2} f^{ABC} \int d^4 x d^4 \theta \left( c'^A V^B c^C + \overline{c}'^A V^B c^C +
  c'^A V^B \overline{c}^C + \overline{c}'^A V^B \overline{c}^C \right)
\eeq

We now quantize the matter sector. From the quadratic part of the action (\ref{eqn:mat-action})
we read the propagators
\bea
&&\langle \bar A^{\hat a}_{\ a}(1) \, A^b_{\ \hat b}(2) \rangle
  = -\frac{1}{\Box} \delta^4(\th_1 - \th_2) \, \delta^{\hat a}_{\ \hat
  b} \, \delta^{\ b}_{a} 
  \\
&&  \langle \bar B^a_{\ \hat a}(1) \, B^{\hat b}_{\ b}(2) \rangle = 
  -\frac{1}{\Box} \delta^4(\th_1 - \th_2) \, \delta^a_{\ b} \, \delta^{\ \hat b}_{\hat a}
  \non \\
&&\langle (\bar Q^1_r)_a(1) \, (Q_1^q)^b(2) \rangle
  = -\frac{1}{\Box} \delta^4(\th_1 - \th_2) \, \delta^{\ b}_{a} \d^{\ q}_{r}
  \non \\
&&\langle (\tilde{Q}_{1,r})_a(1) \, (\bar{\tilde{Q}}^{1,q})^b(2) \rangle
  = -\frac{1}{\Box} \delta^4(\th_1 - \th_2) \, \delta^{\ b}_{a} \d^{\ q}_{r} 
  \qquad \qquad \quad r,q = 1, \cdots , N_f
  \non \\
&&\langle (\bar Q^2_{r'})_{\hat a}(1) \, (Q_2^{q'})^{\hat b}(2) \rangle
  = -\frac{1}{\Box} \delta^4(\th_1 - \th_2) \, \delta^{\ \hat b}_{\hat a} \d^{\ q'}_{r'} 
  \non \\
&&\langle (\tilde{Q}_{2,r'})_{\hat a}(1) \, (\bar{\tilde{Q}}^{2,q'})^{\hat b}(2) \rangle
  = -\frac{1}{\Box} \delta^4(\th_1 - \th_2) \, \delta^{\ \hat b}_{\hat a} \d^{\ q'}_{r'} 
  \qquad \qquad r',q' = 1, \cdots , N'_f
  \non        
\eea
From the expansion of (\ref{eqn:mat-action}) mixed gauge/matter vertices entering two--loop calculations are
\bea
&\action_{\mathrm{mat}}& \rightarrow \intsup \Tr \left( \bar A V A - \bar A
   A \hat V + \bar B \hat V B - \bar B B V \right) 
   \\ 
   && + \intsup \Tr \left( \frac{1}{2} \bar A VVA + \frac{1}{2} \bar
   AA \hat V \hat V - \bar A V A \hat V + \frac{1}{2} \bar B \hat
   V \hat V B + \frac{1}{2} \bar B B VV - \bar B \hat V
   BV \right) 
   \non \\
   && + \intsup \Tr \left( {\bar Q}^1_r V Q^r_1
   - \bar{\tilde{Q}}^{1,r} \tilde Q_{1,r} V + \bar Q^2_{r^\prime}
  {\hat V} Q_2^{r^\prime} - \bar{\tilde Q}^{2, r^\prime} \tilde Q_{2,
  r^\prime} {\hat V} \right)
  \non \\
  && + \intsup \Tr \left( \frac12 {\bar Q}^1_r VV Q^r_1
  + \frac12 \bar{\tilde{Q}}^{1,r} \tilde Q_{1,r} VV + \frac12 \bar Q^2_{r^\prime}
  {\hat V}{\hat V} Q_2^{r^\prime} + \frac12 \bar{\tilde Q}^{2, r^\prime} \tilde Q_{2,
  r^\prime} {\hat V}{\hat V} \right)
  \non
\eea
Pure matter vertices can be read from the superpotential (\ref{eqn:superpot}).

\section{Two--loop renormalization and $\b$--functions} 
 
It is well known that even in the presence of matter chiral superfields the CS actions 
cannot receive loop divergent corrections \cite{Kazakov, KLL}. In fact, gauge invariance
requires $2\pi K_1, 2\pi K_2$ to be integers, so preventing any renormalization except for a 
finite shift. In particular, for the ${\cal N}=2$ case it has been proved \cite{KLL} that even finite
renormalization is absent. 

Divergent contributions are then expected only in the matter sector. Since a non--renormalization theorem
still holds for the superpotential (in ${\cal N}=2$ superspace perturbative calculations one can never
produce local, chiral divergent contributions) divergences arise only in the Kahler sector and lead to 
field functions renormalization.  
 
In odd spacetime dimensions there are no UV divergences at odd loops. Therefore, the first non trivial tests
for the perturbative quantum properties of the theory arise at two loops.

\subsection{One loop results}

We first compute the finite quantum corrections to the scalar and gauge propagators which then enter 
two-loop computations.

\begin{figure}
  \center
  \includegraphics[width=0.35\textwidth]{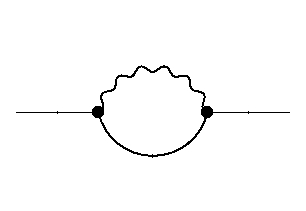}
  \caption{One--loop diagrams for scalar propagators.}
  \label{fig:scalaroneloop}
\end{figure}

The only diagrams contributing to the matter field propagators are the ones given in 
Fig. \ref{fig:scalaroneloop}. It is easy to verify that they vanish for symmetry reasons.

We then move to the gauge propagator. Gauge one-loop self--energy contributions come from diagrams in
Fig. \ref{fig:gaugeoneloop} where chiral, gauge and ghost loops are present. 

\begin{figure}
  \center
  \includegraphics[width=1.0\textwidth]{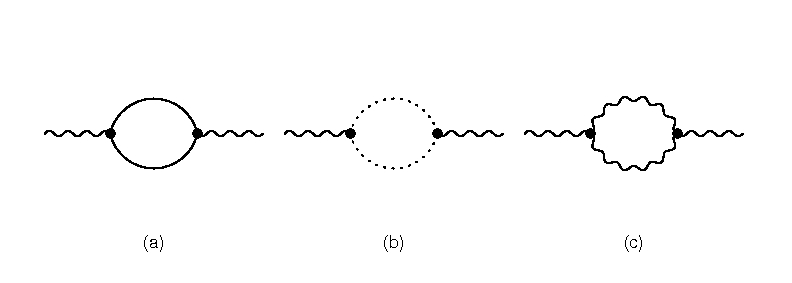}
  \caption{ One--loop diagrams for gauge propagators. }
  \label{fig:gaugeoneloop}
\end{figure}

Performing the calculation in momentum space and using the superspace projectors \cite{superspace}
\beq
  \Pi_0 \equiv -\frac{1}{k^2} \{ D^2, \Db^2 \} (k) \quad , \quad
  \Pi_{1/2} \equiv \frac{1}{k^2} \Db^\a D^2 \Db_\a (k)
\qquad \qquad \Pi_0 + \Pi_{1/2} = 1
\eeq
we find the following finite contributions to the quadratic action for the gauge fields 
\begin{eqnarray}
  \Pi^{(1)}_{gauge}(2a) &=& \frac{1}{8} f^{ABC} f^{A^\prime BC} \intsupk{k} B_0(k)\,
  k^2\, V^A(k)\, \Pi_0 \, V^{A^\prime}(-k) \non \\ 
  \Pi_{gauge}^{(1)}(2b) &=&
  -\frac{1}{8} f^{ABC} f^{A^\prime BC} \intsupk{k} B_0(k)\, k^2\,
  V^A(k) \left( \Pi_0 + \Pi_{1/2} \right) V^{A^\prime}(-k) 
  \non \\ 
  \Pi_{gauge}^{(1)}(2c) &=& \left( M + \frac{N_f}{2} \right) \delta^{AA^\prime} \intsupk{k} B_0(k)\,
  k^2\, V^A(k) \, \Pi_{1/2}\, V^{A^\prime}(-k)
\end{eqnarray}

\bea
  &\hat \Pi_{gauge}^{(1)}(2a) =& \frac{1}{8} \hat f^{ABC} \hat f^{A^\prime
  BC} \intsupk{p} B_0(p)\, p^2\, \hat V^A(p) \, \Pi_0 \, \hat
  V^{A^\prime}(-p) \non \\ 
  &\hat \Pi_{gauge}^{(1)}(2b) =& -\frac{1}{8} \hat f^{ABC} \hat
  f^{A^\prime BC} \intsupk{p} B_0(p)\, p^2\, \hat V^A(p) \left( \Pi_0
  + \Pi_{1/2} \right) \hat V^{A^\prime}(-p) \non \\ 
  &\hat \Pi_{gauge}^{(1)}(2c) =& \left( 
  N + \frac{N_f^\prime}{2} \right) \delta^{AA^\prime} \intsupk{p} B_0(p)\,
  p^2\, \hat V^A(p) \, \Pi_{1/2}\, \hat V^{A^\prime}(-p)
\eea

\bea
  \tilde \Pi_{gauge}^{(1)}(2c) &=& -2 \sqrt{NM} \delta^{A 0} \delta^{A^\prime
  0} \intsupk{p} B_0(p)\, p^2\, V^A(p) \, \Pi_{1/2} \, \hat
  V^{A^\prime}(-p)
\label{mixed}  
\eea
where $B_0(p) = 1/(8|p|)$ is the three dimensional bubble scalar integral (see (\ref{1integral})). 

Summing all the contributions we see that the gauge loop cancels against part of 
the ghost loop as in the 4D ${\cal N}=1$ case \cite{GRS} and we find the known results \cite{GRS, ASW}
\bea
&&  \Pi^{(1)}_{gauge} = \left[ - \frac{1}{8} f^{ABC} f^{A^\prime BC} + \left(
  M+\frac{N_f}{2} \right) \delta^{AA^\prime} \right] \intsupk{p} B_0(p)\,
  p^2\, V^A(p)\, \Pi_{1/2} \, V^{A^\prime}(-p)
\non \\
&&  \hat \Pi^{(1)}_{gauge} = \left[ - \frac{1}{8} f^{ABC} f^{A^\prime BC} + \left(
  N+\frac{N'_f}{2} \right) \delta^{AA^\prime} \right] \intsupk{p} B_0(p)\,
  p^2\, {\hat V}^A(p)\, \Pi_{1/2} \, {\hat V}^{A^\prime}(-p) 
  \non \\
\eea
together with $\tilde \Pi_{gauge}^{(1)} $ in (\ref{mixed}) which mixes the two $U(1)$ gauge sectors.

\subsection{Two-loop results}

We are now ready to evaluate the matter self--energy contributions at two loops. Both for the bifundamental
and the flavor matter the divergent diagrams are given in Fig. \ref{fig:matter2loop}. 

\begin{figure}
  \center
  \includegraphics{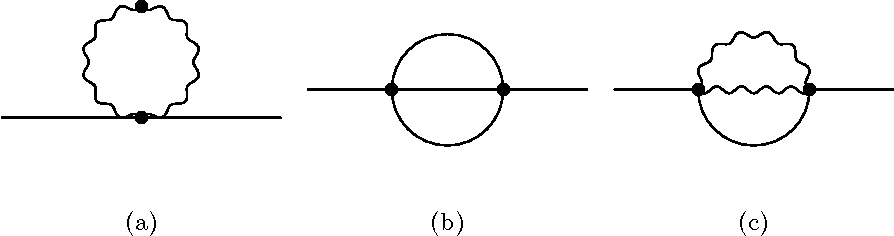}
  \caption{ Two--loop divergent diagrams contributing to the matter propagators.}
  \label{fig:matter2loop}
\end{figure}

Evaluation of each diagram
proceeds in the standard way by first performing D--algebra in order to reduce supergraphs to ordinary 
Feynman graphs and evaluate them in momentum space and dimensional regularization ($d = 3-2\e$).
Separating the contributions of each diagram, the results for the bifundamental matter are
\bea \label{eqn:bifundamental}
  &&\Pi^{(2)}_{bif}(3a) = -\left[ \frac{1}{K_1^2} \left( 2NM + N N_f
  - \frac{1}{2}\left( N^2 -1\right) \right)
  + \right.
  \non \\
  &&~~~~~~~~~\left.+\frac{1}{K_2^2} \left( 2NM + M N_f^\prime
  -\frac{1}{2} \left(M^2 - 1\right)\right) +\frac{4}{K_1\,
  K_2} \right] \, F(0) \; \Tr \left( \bar A_i A^i + \bar B^i B_i \right) 
  \non \\ 
  \non \\
  &&\Pi^{(2)}_{bif}(3b) = \Big[ 4|h_1|^2 (MN +1) + (|h_3|^2+|h_4|^2) MN + (h_3 \barh_4 + h_4 \barh_3)
  \non \\
  &&~~ \quad \quad \quad +( |\a_1|^2 N N_f + |\a_3|^2 M
  N_f^\prime) \Big] \, F(p) \; \Tr \left( \bar A_1 A^1 + \bar B^1 B_1\right) 
  \non \\ 
  && ~~~ \quad  + \Big[ 4|h_2|^2 (MN +1) + (|h_3|^2+|h_4|^2) MN
  + (h_3 \barh_4 + h_4 \barh_3) 
  \non \\
  &&~~ \quad \quad \quad +( |\a_2|^2 N N_f + |\a_4|^2 M N_f^\prime ) \Big] \, F(p) \;
  \Tr \left( \bar A_2 A^2 + \bar B^2 B_2 \right) 
  \non \\
  \non \\
  &&\Pi^{(2)}_{bif}(3c) = -\frac{1}{2} \left[ \frac{N^2+1}{K_1^2}
  + \frac{M^2+1}{K_2^2} + \frac{4 NM}{K_1\, K_2} \right] \, F(p) \;
  \Tr \left( \bar A_i A^i + \bar B^i B_i \right)
\eea
where $F(p)$ is the two--loop self--energy integral given in (\ref{integral}).
 
\vskip 10pt 
\noindent
Analogously, for fundamental matter we find
\bea \label{eqn:fundamental}
  &&\Pi^{(2)}_{fund1}(3a) =-  \frac{1}{K_1^2} \left( 2NM + N N_f
  - \frac{1}{2}\left( N^2 -1\right) \right) \, F(0) \,
  \Tr \left( \bar Q^1 Q_1 + \bar{\tilde Q}^1 \tilde Q_1 \right) 
  \non \\
  &&\Pi^{(2)}_{fund2}(3a) = -\frac{1}{K_2^2} \left( 2NM+ M N_f^\prime 
  -\frac{1}{2} \left(M^2 - 1\right)\right)  \, F(0) \,  
  \Tr \left( \bar Q^2 Q_2 + \bar{\tilde Q}^2 \tilde Q_2 \right) 
  \non \\
  \non \\
  &&\Pi^{(2)}_{fund1}(3b) = ~~\Big[ 4|\l_1|^2 (N N_f +1) + |\l_3|^2 MN_f^{\prime} 
  \non \\
  && \qquad \qquad \qquad \qquad \qquad \qquad + 
  \left(|\a_1|^2+|\a_2|^2\right) MN\Big] \, F(p) \,  
  \Tr \left( \bar Q^1 Q_1 + \bar{\tilde Q}^1 \tilde Q_1\right)
  \non \\
  && \Pi^{(2)}_{fund2}(3b) = \Big[ 4|\l_2|^2 (M N_f^{\prime} +1) + |\l_3|^2 NN_f 
  \non \\
  && \qquad \qquad \qquad \qquad \qquad \qquad + 
  \left(|\a_3|^2+|\a_4|^2\right) NM \Big] \, F(p) \,  
  \Tr \left( \bar Q^2 Q_2 + \bar{\tilde Q}^2 \tilde Q_2\right) 
  \non \\
  \non \\
  &&\Pi^{(2)}_{fund1}(3c) =  -\frac{N^2+1}{2K_1^2} \, F(p) \, 
  \Tr \left( \bar Q^1 Q_1 + \bar{\tilde Q}^1 \tilde Q_1 \right)
  \non \\
  && \Pi^{(2)}_{fund2}(3c) = - \frac{M^2+1}{2K_2^2} \, F(p) \,   
  \Tr \left( \bar Q^2 Q_2 + \bar{\tilde Q}^2 \tilde Q_2 \right)
\eea
where $F(p)$ is still given in (\ref{integral}).

\vskip 15pt
We now proceed to the renormalization of the theory. We define
renormalized fields as
\beq
\Phi = Z_{\Phi}^{-\frac12} \Phi_B    \qquad , \qquad \Phib = \bar{Z}_{\Phib}^{-\frac12} \Phib_B
\eeq
where $\Phi$ stands for any chiral field of the theory, and coupling constants as
\bea
\begin{array}{lcllcllcllcl}
  h_j &=& \mu^{-2\e} Z_{h_j}^{-1} h_{j \,
  B} \qquad   &\bar h_j &=& \mu^{-2\e} Z_{\bar
  h_j}^{-1} \bar h_{j \, B} \non
\\
  \l_j &=& \mu^{-2\e} Z_{\l_j}^{-1} \l_{j\,
  B} \qquad  &\bar \l_j &=& \mu^{-2\e}
  Z_{\bar \l_j}^{-1} \bar \l_{j \, B} \non
\\
  \a_j &=& \mu^{-2\e} Z_{\a_j}^{-1} \a_{j \,
  B} \qquad  &\bar \a_j &=& \mu^{-2\e}
  Z_{\bar \a_j}^{-1} \bar \a_{j\,B} \non
\end{array}
\eea
together with $K_1 = \mu^{2\e}K_{1\,B}, K_2 = \mu^{2\e}K_{2\,B}$.
Powers of the renormalization mass $\mu$ have been introduced in
order to deal with dimensionless renormalized couplings. 

In order to cancel the divergences in (\ref{eqn:bifundamental}) and
(\ref{eqn:fundamental}) we choose
\bea
\label{eqn:ZA1}
  &&Z_{A^1} = Z_{\bar A_1} = Z_{B_1} = Z_{\bar B^1} = 1 - 
  \\ 
  &&  \frac{1}{64\pi^2} \Big[ -\frac{2NM + N N_f + 1}{K_1^2} - \frac{2NM
  + M N^\prime_f + 1}{K_2^2} - \frac{2NM + 4}{K_1 K_2}  
  \non \\ 
  && +4|h_1|^2 (MN +1) + (|h_3|^2+|h_4|^2) MN +
  (h_3 \barh_4 + h_4 \barh_3) + ( |\a_1|^2 N N_f + |\a_3|^2 M
  N_f^\prime) \Big] \frac{1}{\e} 
\non \\ 
\non \\
  &&Z_{A^2} = Z_{\bar A_2} = Z_{B_2} = Z_{\bar B^2} = 1- 
  \non \\
&&  \frac{1}{64\pi^2} \left[ -\frac{2NM + N N_f + 1}{K_1^2} - \frac{2NM
  + M N^\prime_f + 1}{K_2^2} - \frac{2NM + 4}{K_1 K_2} \right. 
  \non \\
  && \left. +4|h_2|^2 (MN +1) + (|h_3|^2+|h_4|^2) MN +
  (h_3 \barh_4 + h_4 \barh_3) + ( |\a_2|^2 N N_f + |\a_4|^2 M
  N_f^\prime)  \right] \frac{1}{\e} \non
\non \\
\non \\
  &&Z_{Q_1} = Z_{\bar Q^1} = Z_{\tilde Q_1} = Z_{\bar{\tilde Q}^1} = 1 -
  \non \\
&&  \frac{1}{64\pi^2} \left[ -\frac{2NM + N N_f +
  1}{K_1^2} + 4|\l_1|^2\, (N N_f +1) + |\l_3|^2 MN^\prime_f + ( |\a_1|^2 + |\a_2|^2 ) M
  N \right] \frac{1}{\e} 
\non \\
\non \\
  &&Z_{Q_2} = Z_{\bar Q^2} = Z_{\tilde Q_2} = Z_{\bar{\tilde Q}^2} = 1 -
  \non \\
&&  \frac{1}{64\pi^2} \left[ -\frac{2NM + M N^\prime_f +
  1}{K_2^2} + 4|\l_2|^2\, (M
  N^\prime_f+1) + |\l_3|^2 N N_f + ( |\a_3|^2 + |\a_4|^2 ) M
  N \right] \frac{1}{\e} 
\non 
\eea
Thanks to the non-renormalization theorem for the superpotential, the renormalization of the 
couplings is a consequence of the field renormalization. In particular, we set
\bea \label{eqn:Zlambda}
  Z_{\n_j} = \prod_{\Phi_i} Z^{-\frac12}_{\Phi_i}
\eea
where $\n_j$ stands for any coupling of the theory and the sum is extended to all the $\Phi_i$ fields 
coupled by $\n_j$.
 
The anomalous dimensions and the beta-functions are given by the general prescription
\bea
  \g_{\Phi_j} &\equiv& \frac12 \frac{\pa \log{Z_{\Phi_j}}}{\pa \log{\mu}}
  = - \frac12 \sum_i d_i \, \n_i \, \frac{\pa
  Z^{(1)}_{\Phi_j}}{\pa \n_i} \label{eqn:anomdim}
\\
  \b_{\n_j} &=& -d_j \, \n^{(1)}_j
  + \sum_i \left( d_i \, \n_i \, \frac{\pa \n^{(1)}_j}{\pa \n_i} \right) 
  = \n_j(\mu) \sum_i \g_i \label{eqn:beta}
\eea
where $d_j$ is the bare dimension of the $\n_j$--coupling and $Z^{(1)}_{\Phi_j}$ is the coefficient of the $1/\e$ pole in
$Z_{\Phi_j}$. The last equality in (\ref{eqn:beta}) follows from
(\ref{eqn:anomdim}) and (\ref{eqn:Zlambda}). 

Reading the single pole coefficient $Z^{(1)}_{\Phi_j}$ in
eqs. (\ref{eqn:ZA1}) we finally obtain
\bea
  &&\g_{A^1} = \g_{B_1} = \frac{1}{32\pi^2} \Big[ -\frac{2NM + N N_f +
  1}{K_1^2} - \frac{2NM + M N^\prime_f + 1}{K_2^2} - \frac{2NM +
  4}{K_1 K_2} 
  \non \\ 
  &&~~~~~~~~~~~~~~~~ \qquad \quad + 4|h_1|^2 (MN +1)
  + (|h_3|^2+|h_4|^2) MN + (h_3 \barh_4 + h_4 \barh_3) 
  \non \\ 
  &&~~~~~~~~~~~~~~~~\qquad \quad + ( |\a_1|^2 N N_f +
  |\a_3|^2 M N_f^\prime) \Big] 
  \non \\
  &&\g_{A^2} = \g_{B_2} = \frac{1}{32\pi^2} \Big[ -\frac{2NM + N N_f
  + 1}{K_1^2} - \frac{2NM + M N^\prime_f + 1}{K_2^2} - \frac{2NM +
  4}{K_1 K_2} 
  \non \\ 
  &&~~~~~~~~~~~~~~~~ \qquad \quad + 4|h_2|^2 (MN +1)
  + (|h_3|^2+|h_4|^2) MN + (h_3 \barh_4 + h_4 \barh_3)
  \non \\ 
  &&~~~~~~~~~~~~~~~~ \qquad \quad + ( |\a_2|^2 N N_f+
  |\a_4|^2 M N_f^\prime)  \Big] 
  \non \\
  &&\g_{Q_1} = \g_{\tilde{Q}_1} = \frac{1}{32\pi^2} \Big[ -\frac{2NM +
  N N_f + 1}{K_1^2} 
  \non \\
  &&~~~~~~~~~~~~~~~~ \qquad \quad  + 4|\l_1|^2\, (N N_f+1) + |\l_3|^2
  MN^\prime_f + ( |\a_1|^2 + |\a_2|^2 ) M N \Big] 
  \non \\
  &&\g_{Q_2} = \g_{\tilde{Q}_2} = \frac{1}{32\pi^2} \Big[ -\frac{2NM +
  M N^\prime_f + 1}{K_2^2}  
  \non \\
  &&~~~~~~~~~~~~~~~~ \qquad \quad  + 4|\l_2|^2\, (M N^\prime_f+1) + |\l_3|^2 N
  N_f + ( |\a_3|^2 + |\a_4|^2 ) M N \Big] \non
\\
\label{gamma}
\eea 
whereas the corresponding beta--functions are given by
\bea
&& \b_{h_1} = 4 h_1 \g_{A^1} \qquad \qquad \qquad \quad \b_{h_2} = 4 h_2 \g_{A^2} \non \\
&&   \b_{h_3} = 2h_3 ( \g_{A^1} + \g_{A^2} ) \qquad \qquad \b_{h_4} = 2h_4 ( \g_{A^1} + \g_{A^2} ) \non \\
&& \b_{\l_1} = 4 \l_1 \g_{Q_1} \qquad \qquad \qquad \quad  \b_{\l_2} = 4 \l_2 \g_{Q_2} \non \\ 
&& \qquad \qquad \qquad \b_{\l_3} = 2\l_3 ( \g_{Q_1} + \g_{Q_2} ) \non \\
&&  \b_{\a_1} = 2\a_1 ( \g_{A_1} + \g_{Q_1} ) \qquad \qquad   \b_{\a_2} = 2\a_2 ( \g_{A_2} + \g_{Q_1} ) 
\non \\
&&  \b_{\a_3} = 2\a_3 ( \g_{A_1} + \g_{Q_2} )\qquad \qquad  \b_{\a_4} = 2\a_4 ( \g_{A_2} + \g_{Q_2} )
\label{beta}
\eea

\section{The spectrum of fixed points}

In this Section we study solutions to the equations $\b_{\n_j} = 0$ where the beta--functions are
given in (\ref{beta}). We consider separately the cases with and without flavor matter.

\subsection{Theories without flavors}

\noindent
We begin by considering the class of theories without flavors. In eqs. (\ref{gamma}) we set
$N_f = N'_f = 0$, $\l_j = \a_j = 0$ and solve the equations
\bea
&& \b_{h_1} = 4 h_1 \g_{A^1} =0 \qquad \qquad \qquad \b_{h_2} = 4 h_2 \g_{A^2} =0 \non \\
&&   \b_{h_3} =2 h_3 ( \g_{A^1} + \g_{A^2} ) =0 \qquad \quad \b_{h_4} = 2h_4 ( \g_{A^1} + \g_{A^2} ) =0 
\label{nullbeta}
\eea
When $h_j \neq 0$ for any $j$ the conditions (\ref{nullbeta}) are equivalent to $\g_{A^1} =
\g_{A^2} =0$, that is no UV divergences appear at two--loops. 
On the other hand, if we restrict to 
the surface $h_1=h_2=0$, the beta--functions are zero when $\g_{A^1} + \g_{A^2}  =0$, which in principle 
would not require the anomalous dimensions to vanish. However, it is easy to see from (\ref{gamma}) that 
for $h_1=h_2=0$ we have $\g_{A^1} = \g_{A^2}$ and again $\b_{h_3} = \b_{h_4}=0$ imply the vanishing of
all the anomalous dimensions. Therefore, at two loops
the request for vanishing beta--functions is equivalent to the request of finiteness. 

We first study the class of theories with $h_1=h_2=0$. In this case we find convenient to redefine 
the couplings as \cite{us}
\beq
y_1 = h_3 + h_4 \qquad , \qquad y_2 = h_3 - h_4 
\label{y}
\eeq
In fact, writing the superpotential in terms of the new couplings  
\beq
\int d^4x d^2 \th \, \left[ \frac{y_1}{2} \Tr(A^1 B_1 A^2 B_2 + A^2 B_1 A^1 B_2) \, + \, 
\frac{y_2}{4} \e_{ij} \e^{kl} \Tr( A^i B_k A^j B_l) \right]
\eeq
it is easy to see that $y_1$ is associated to a $SU(2)_A \times SU(2)_B$ breaking perturbation, whereas
$y_2$ is symmetry preserving. 

For real couplings, the anomalous dimensions vanish when 
\beq
y_1^2 (MN+1) + y_2^2 (MN-1) = 
2(2MN + 1)\left(\frac{1}{K_1^2} + \frac{1}{K_2^2}\right) + 2\frac{2MN + 4}{K_1 K_2} 
\label{nullbeta2}
\eeq
This describes an ellipse in the parameter space. For $K_{1,2}$ sufficiently large it is very closed to the origin 
and solutions fall in the perturbative regime. The ellipse degenerates to a circle in the large $M,N$ limit. 
Fixed points corresponding to $y_1 \neq 0$ ($h_4 \neq - h_3$) describe ${\cal N} = 2$ superconformal theories 
with $U(1)_A \times U(1)_B$ global symmetry (\ref{U(1)}). 

A more symmetric conformal point is obtained by solving (\ref{nullbeta2}) under the condition $y_1=0$. 
The solution 
\beq
h_3 = - h_4 = \sqrt{ \frac{2 MN+1}{2(MN-1)} \left( \frac{1}{K_1^2}+\frac{1}{K_2^2} \right) 
+ \frac{MN+2}{MN-1} \, \frac{1}{K_1 K_2} }
\label{SO(4)}
\eeq
corresponds to a superconformal theory with $SU(2)_A \times SU(2)_B$ global symmetry. This is the 
theory conjectured in \cite{GT}. 
When $K_1 = -K_2 \equiv K$ it reduces to $h_3=-h_4 = 1/K$ and we recover the ${\cal N}=6$ ABJ model 
\cite{ABJ} and, for $N=M$, the ABJM one \cite{ABJM}. 

More generally, we study fixed points with $h_j \neq 0$ for any $j$. In this case we have 
two equations, $\g_{A^1} = \g_{A^2}=0$, for four unknowns. The spectrum of fixed points then spans a 
two dimensional surface which for real couplings is given by 
\bea
h_1^2 = h_2^2 &=& \frac{1}{4(MN+1)} \Big[ (2 MN+1) \left( \frac{1}{K_1^2}+\frac{1}{K_2^2} \right)
+ 2\frac{MN+2}{K_1 K_2} - MN(h_3^2 + h_4^2) - 2h_3h_4  \Big]
\non \\
&&~~
\label{surface}
\eea
This equation describes an ellipsoid in the four dimensional $h$--space as given in Fig. 4,
localized in the subspace $h_1=h_2$ (or equivalently $h_1= -h_2$). 
A particular point on this surface corresponds to $h_3= 1/K_1$ and $h_4= 1/K_2$ with, 
consequently, $h_1 = h_2 = \frac12 ( \frac{1}{K_1} + \frac{1}{K_2})$. 
This is the ${\cal N}=3$ superconformal theory discussed in \cite{GT} \footnote{Finiteness properties of ${\cal N}=3$ 
CS--matter theories have also been discussed in \cite{BILPSZ} within the ${\cal N}=3$ harmonic superspace setup.}.

The locus $h_1=h_2=0$, $h_3 = -h_4$ of this surface is the ${\cal N}=2$, $SU(2)_A \times SU(2)_B$ invariant 
superconformal theory (\ref{SO(4)}). Therefore, the ${\cal N}=3$ and the 
${\cal N}=2$, $SU(2)_A \times SU(2)_B$ superconformal points are continuously connected by the surface 
(\ref{surface}). 

We can select a particular line of fixed points interpolating between the two theories, by setting 
\beq
h_1 = h_2 = \frac12 (h_3 + h_4) 
\eeq
and, consequently 
\beq
h_3^2 + h_4^2 + 2 \frac{MN+2}{2MN+1} \, h_3 h_4  =  \frac{1}{K_1^2} + \frac{1}{K_2^2}
+ 2 \frac{MN+2}{K_1 K_2\, (2MN+1)}
\label{line}
\eeq
These are $SU(2)$ invariant, ${\cal N}=2$ superconformal theories with 
superpotential (\ref{potGT}). The existence of a line of $SU(2)$ invariant fixed points interpolating between the two
theories was already conjectured in \cite{GT}. 

\begin{figure}
  \center
  \includegraphics[scale=0.5]{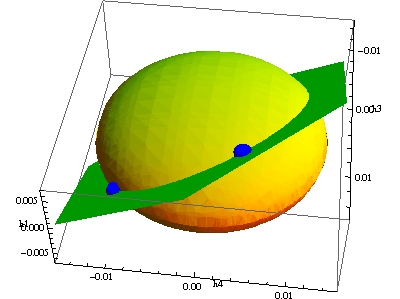}
  \caption{The exactly marginal surface of fixed points in the space of $h_i$
  couplings, restricted to the subspace $h_1=h_2$. The parameters have been chosen as $K_1 = 150, K_2=237, 
  N=43, M=30$. The dots denote the ${\cal N}=3$ and the ${\cal N}=2$,
  $SU(2)_A \times SU(2)_B$ fixed points belonging to the
  ellipsoid. The plane represents the class of theories (\ref{potGT})
  with $SU(2)$ global symmetry and its intersection with the ellipsoid
  is the line described by (\ref{line}).}
  \label{fig:ellipsoid}
\end{figure}

\vskip 10pt

So far we have considered real solutions to the equations $\b_{\nu_j}=0$. We now discuss the case
of complex couplings focusing in particular on the so--called $\b$--deformations. 

In the class of theories with $h_1=h_2=0$ we look for solutions of the form
\beq
h_3 = h e^{i\pi \b} \qquad , \qquad h_4 = -h e^{-i\pi \b} 
\label{betadef}
\eeq
which implies $y_1 = 2h \sin{\pi \b}, \, y_2 = 2h \cos{\pi \b}$ in (\ref{y}). The condition for vanishing beta--functions then reads
\beq
h^2 MN   - h^2 \cos{2\pi \b} =  \frac12 (2MN+1)\left( \frac{1}{K_1^2} + \frac{1}{K_2^2} \right) + \frac{MN +2}{K_1 K_2} 
\label{deformations}
\eeq
This describes a line of fixed points which correspond to superconformal beta--deformations of the
$SU(2)_A \times SU(2)_B$ invariant theory (\ref{SO(4)}). For $\b \neq 0$ the global symmetry is broken to $U(1)_A \times
U(1)_B$ in (\ref{U(1)}) and the deformed theory is only ${\cal N}=2$ supersymmetric. 
In particular, setting $K_1 = -K_2$ we obtain the $\b$--deformed ABJM/ABJ theories studied in \cite{Imeroni}.

In the large $M,N$ limit the $\b$--dependence of equation (\ref{deformations}) disappears, consistently with the fact
that in planar Feynman diagrams  the effects of the deformation are invisible \cite{MPSZ}. In this limit the condition for superconformal invariance reads 
\beq
h^2 = \frac{1}{K_1^2} + \frac{1}{K_2^2} + \frac{1}{K_1 K_2} 
\eeq
which reduces to $h=1/K$ for opposite CS levels. 

The analysis of $\b$--deformations can be extended to theories with $h_1, h_2 \neq 0$. Since they enter the anomalous 
dimensions  only through $|h_1|^2$ and $|h_2|^2$  we can take them to be arbitrarily complex and still make
the ansatz (\ref{betadef}) for $h_3, h_4$. 
The surface of fixed points is then given by
\bea
|h_1|^2 = |h_2|^2 &=& \frac{1}{4(MN+1)} \Big[ (2 MN+1) \left( \frac{1}{K_1^2}+\frac{1}{K_2^2} \right)
+ 2\frac{MN+2}{K_1 K_2} 
\non \\
&& \qquad \qquad \qquad - 2 h^2 MN + 2h^2 \cos{2\pi \b}  \Big]
\eea
and describes superconformal $\b$--deformations of ${\cal N}=2$ invariant theories. 

The results of this Section agree with the ones in \cite{ASW} obtained by using the three--algebra formalism. 

\subsection{Theories with flavors}

As in the previous case, when all the couplings
are non-vanishing, the request for zero beta--functions implies the finiteness conditions $\g_{\Phi_i}=0$. 
These provide four constraints on a set of eleven unknowns (see eqs. (\ref{gamma})).
Therefore, in the space of the coupling constants the spectrum of fixed points spans a seven dimensional 
hypersurface given by the equations 
\bea \label{eqn:finiteness}
&&
  |\a_2|^2 = \frac{1}{ N N_f K_1^2 K_2^2 }\Big\{K_2^2 \left(2NM + N N_f +
  1\right) + K_1^2 \left(2NM + M N^\prime_f + 1\right) 
  \non \\
  && \qquad \qquad \qquad \qquad \quad  + 2 K_1 K_2 \left(NM + 2\right) -  4|h_2|^2 \, K_1^2 K_2^2 
  \,(MN +1)  
\non \\
&& \qquad \qquad \qquad \qquad \quad  \frac{}{}  
 - K_1^2 K_2^2 \left[ (|h_3|^2+|h_4|^2) MN + (h_3 \barh_4 + h_4 \barh_3) + |\a_4|^2 M N_f^\prime 
 \right]\Big\} \non
\\
\non 
\eea
\bea
&& |\a_3|^2 = \frac{1}{ M N_f^\prime K_1^2 K_2^2 } \Big\{ K_2^2 \left(2NM + N N_f + 
   1\right) + K_1^2 \left(2NM + M N^\prime_f + 1\right) 
  \non \\
  && \qquad \qquad \qquad \qquad \quad  + 2 K_1 K_2 \left(NM + 2\right) -   4|h_1|^2 \, 
  K_1^2K_2^2 \, (MN +1)
\non \\
&& \qquad \qquad \qquad \qquad \quad  \frac{}{}  
   - K_1^2K_2^2 \left[ (|h_3|^2+|h_4|^2) MN + (h_3 \barh_4 + h_4 \barh_3) + |\a_1|^2 N N_f 
   \right]\Big\} \non
\\  
\non 
\eea
\bea
&&  |\l_1|^2 = \frac{1}{4 (N N_f+1)  K_1^2} \Big\{ 2NM +
  N N_f + 1  - K_1^2\left[ |\l_3|^2
  MN^\prime_f + ( |\a_1|^2 + |\a_2|^2 ) M N \right] \Big\}  \non
\\
\non \\
&&
  |\l_2|^2 = \frac{1}{4 (M N^\prime_f+1)  K_2^2} \Big\{ 2NM +
  M N^\prime_f + 1  - K_2^2\left[ |\l_3|^2
  NN_f + ( |\a_3|^2 + |\a_4|^2 ) M N \right] \Big\} 
  \non
\\  
\label{nullbeta3}    
\ena
When $K_1 = - K_2 \equiv K$ a particular point on this surface corresponds to
\bea
&& h_1 = h_2 = 0 \qquad ~~~ ,\qquad h_3 = - h_4 = \frac{1}{K} 
\non \\
&& \l_1 = - \l_2 = \frac{1}{2K} ~~~, \qquad \l_3 = 0
\non \\
&& \a_1 = \a_2 = \frac{1}{K} \qquad ,\qquad  \a_3 = \a_4 = -\frac{1}{K}
\label{ABJMflavor}
\eea
and describes the ${\cal N}=3$ ABJ/ABJM models with flavor matter \cite{HK, GJ, HLT}. 

More generally, allowing $K_2 \neq - K_1$ we find the fixed point
\bea
&& h_1 = h_2 = \frac12 \left( \frac{1}{K_1} +  \frac{1}{K_2} \right)
\quad ,\quad h_3 = \frac{1}{K_1} \quad , \quad  h_4 = \frac{1}{K_2} 
\non \\
&& \l_1 = \frac{1}{2K_1} \quad , \quad  \l_2 = \frac{1}{2K_2} ~~, \qquad \l_3 = 0
\non \\
&& \a_1 = \a_2 = \frac{1}{K_1} \qquad ,\qquad  \a_3 = \a_4 = \frac{1}{K_2}
\label{new}
\eea
which corresponds to a superconformal theory obtained from the ${\cal N} = 3$ theory 
of \cite{GT} by the addition of flavor matter \cite{GJ}. The superpotential 
\bea
{\cal S}_{\mathrm{pot}} &=& \int d^3x\,d^2\theta\:
   \Tr  \left\{ \frac12 \left( \frac{1}{K_1} +  \frac{1}{K_2} \right) \left[ (A^1 B_1)^2 + (A^2 B_2)^2 
   \right] \right. 
   \\
   && \left.  +  \frac{1}{K_1} (A^1 B_1 A^2 B_2) + \frac{1}{K_2} (A^2 B_1 A^1 B_2) 
   + \frac{1}{2K_1} (Q_1 \tilde Q_1)^2 + \frac{1}{2K_2}
  (Q_2 \tilde Q_2)^2  \right.
  \non \\ 
  &&
  \left. + \frac{1}{K_1} \left[ \tilde Q_1 A^i B_i Q_1 \right]
  + \frac{1}{K_2} \left[ \tilde Q_2 B_i A^i Q_2 \right] \right\}  \, + \, h.c.
\non 
\eea
can be thought of as arising from the action
\bea
{\cal S} &=& {\cal S}_{\mathrm{CS}} + {\cal S}_{\mathrm{mat}} 
\non \\
&+&   \int d^3x\,d^2\theta\:
   \left[ -\frac{K_1}{2} \Tr (\Phi_1^2) + \Tr (B_i \Phi_1 A^i) + \Tr (\tilde Q_1 \Phi_1 Q_1)
   \right]
  \non \\ 
&+&   \int d^3x\,d^2\theta\:
   \left[ -\frac{K_2}{2} \Tr (\Phi_2^2) + \Tr (A^i \Phi_2 B_i) + \Tr (\tilde Q_2 \Phi_2 Q_2)
   \right] \, + \, h.c.
\eea
after integration on the $\Phi_1, \Phi_2$ chiral superfields belonging to the adjoint representations of 
the two gauge groups and giving the ${\cal N}=4$ completion of the vector multiplet. Therefore, as in the
unflavored case, the theory exhibits ${\cal N}=3$ supersymmetry with the couples 
$(A,B^{\dag})_i$, $(Q, \tilde{Q}^{\dag})^r_1$ and $(Q, \tilde{Q}^{\dag})^{r'}_2$ realizing $(2+N_f+N'_f)$
${\cal N}=4$ hypermultiplets (The CS terms break ${\cal N}=4$ to ${\cal N} = 3$). 

As already discussed, in the absence of flavors the ${\cal N}=3$ superconformal theory is connected 
by the line of fixed points (\ref{line}) to a ${\cal N}=2$, $SU(2)_A \times SU(2)_B$ invariant theory. 
We now investigate whether a similar pattern arises even when flavors are present. 

To this end, we first choose
\beq
h_1 = h_2 = \frac12 (h_3 +h_4) \qquad , \qquad  \a_1 = \a_2 \quad, \quad  \a_3 = \a_4
\label{c1c2}  
\eeq
with $\l_j$ arbitrary. This describes a set of ${\cal N}=2$ theories with global $SU(2)$ invariance in the bifundamental sector. 
 
Solving the equations $\b_{\n_j}=0$  for real couplings we find a whole line 
of $SU(2)_A \times SU(2)_B$ invariant fixed points parametrized by the unconstrained coupling $\l_3$ 
\bea
\a_1 = \a_3 &=& 0 
\non \\
h_3 = - h_4 &=&  \sqrt{ \frac{(2 N M + M N^\prime_f + 1) K_1^2
    + 2 (M N+2) K_1 K_2 + (2 M N + N N_f + 1) K_2^2}{2(M N-1) K_1^2
    K_2^2} }
\non \\
\l_1^2 &=&  \frac{ 2 M N+N N_f + 1-K_1^2 M N^\prime_f \l_3^2}{4K_1^2 (N N_f + 1)}
\non \\
\l_2^2 &=& \frac{ 2 M N + M N^\prime_f + 1-K_2^2 N N_f \l_3^2}{4K_2^2 (M N^\prime_f + 1)}
\label{SO4}
\eea
A four dimensional hypersurface of ${\cal N}=2$ fixed points given by
\bea 
\a_1^2 &=& \frac{1}{2 M N K_1^2} \, \Big[ -4 K_1^2 (N N_f + 1)
\l_1^2 + N N_f + 2MN +1 - M N^\prime_f \l_3^2 K_1^2 \Big] 
\\ 
\a_3^2 &=& \frac{1}{2 M N K_2^2} \Big[ -4 K_2^2 (M N^\prime_f + 1) \l_2^2 + M
N^\prime_f + 2 M N + 1 - N N_f \l_3^2 K_2^2 \Big] 
\non \\ 
h_3 &=& -\frac{1}{2 M N+1}\Bigg\{ (MN+2)h_4 \pm \Bigg[(2 M N+1) \Bigg( M
N_f^\prime \left(-\a_3^2+4 \lambda_2^2+\lambda_3^2\right) \nonumber \\
&&~~~~~~~~~~~~~~~~~~ +2 N M (\a_1^2+\a_3^2 +1) + N N_f \left(-\a_1^2+4
\lambda_1^2+\lambda_3^2\right) \nonumber \\ &&~~~~~~~~~~~~~~~~~~ + 4
\left(\lambda_1^2+\lambda_2^2\right)+\frac{4}{K_1K_2} \Bigg) -3 h_4^2
\left(M^2 N^2-1\right) \Bigg]^{1/2} \Bigg\} 
\non 
\eea
connects the line of ${\cal N}=2$, $SU(2)_A \times SU(2)_B$ invariant theories (\ref{SO4}) to the ${\cal N}=3$ theory 
(\ref{new}). This is the analogous of the fixed line (\ref{line}) found in the unflavored theories.

\vskip 10pt
Before closing this Section we address the question of superconformal invariance versus finiteness for theories with 
flavor matter. In the bifundamental sector, the only possibility to have vanishing beta--functions 
without vanishing anomalous dimensions is by setting $h_1 = h_2 = 0$. When flavor matter is present,
this does not necessarily imply $\g_{A^1} = \g_{A^2}$, so we can solve for $\b_{h_3, h_4} = 
\g_{A^1} + \g_{A^2}=0$ without requiring the two $\g$'s to vanish separately. Once these equations 
have been solved in the bifundamental sector, in the flavor sector we choose $\l_1=
\l_2 =0$ and $\a_1 = \a_4 = 0$ (or equivalently, $\a_2 = \a_3 = 0$) in order to avoid $\g_{Q_1} = 
\g_{Q_2} = 0$. We are then left with five couplings subject to the three equations $\g_{A^1} + \g_{A^2}=0$,
 $\g_{A^1} + \g_{Q_2}=0$ and $\g_{A^2} + \g_{Q_1} = 0$. Solutions correspond to superconformal but {\em not finite} 
theories.  We note that this is true as long as we work with $M,N$ finite. In the large $M,N$ limit with 
$N_f, N^\prime_f \ll M,N$ we are back to $\g_{A^1} = \g_{A^2}$, as flavor contributions are subleading. In this 
case superconformal invariance requires finiteness.

\section{Infrared stability}

We now study the RG flows around the fixed points of main interest in order to establish whether 
they are IR attractors or repulsors. In particular, we concentrate on the ABJ/ABJM theories, ${\cal N}=3$ and 
$SU(2)_A \times SU(2)_B$ ${\cal N}=2$ superconformal points, in all cases with and without flavors.  

The behavior of the system around a given fixed point $\n_0$ is determined by studying the stability matrix
\beq
{\cal M}_{ij} \equiv \frac{d \b_i}{d \n_j}(\n_0)  
\label{stability}
\eeq
Diagonalizing ${\cal M}$, positive eigenvalues correspond to directions of increasing
$\b$--functions, whereas negative eigenvalues give decreasing betas. It follows that the fixed point is IR 
stable if ${\cal M}$ has all positive eigenvalues, whereas negative eigenvalues represent directions where 
a classically marginal operator becomes relevant. 

If null eigenvalues are present we need compute derivatives of the stability matrix
along the directions individuated by the corresponding eigenvectors. If along a null direction the second 
derivative of the beta--function is different from zero, then the function has a parabolic behavior and 
the system is unstable. 

We apply these criteria to the two--loop beta--functions (\ref{beta}).

\subsection{Theories without flavors}

We begin with the ${\cal N}=2$ theories without flavor discussed in Section 4.1.
As shown, the nontrivial fixed points lie on a two dimensional ellipsoid and particular points on it are the ${\cal N}=3$ and 
the ${\cal N}=2$ $SU(2)_A\times SU(2)_B$ invariant theories. Since the ellipsoid is localized in the subspaces $h_1= \pm h_2$ 
we restrict our discussion to the $h_1 = h_2$ case. 

When $K_1 = - K_2 \equiv K$, the set of theories with $h_1 = h_2 =0$ has been already studied in \cite{us}. 
In this case the ellipsoid reduces to an ellipse in the $(y_1,y_2)$ plane with $y_1$ and $y_2$ defined in eq. (\ref{y}). 
It has been shown that at the order we are working the RG trajectories are straight lines passing through the origin
and intersecting the ellipse. Infrared flows point towards the ellipse so proving that the whole line of 
fixed points is IR stable. However, every single point has only one direction of stability which corresponds to the RG trajectory
passing through it.  When perturbed along any other direction the system flows to a different fixed point on the curve. 
In the ABJ/ABJM case the direction of stability is described by $SU(2)_A \times SU(2)_B$ preserving perturbations.
 
This can be understood by computing the stability matrix at $h_1 = h_2 = 0, \, h_3 = - h_4 = 1/K$ and 
diagonalizing it. We find 
that mutual orthogonal directions are $(h_1= h_2, y_1, y_2)$ and the corresponding eigenvalues are
\beq
{{\cal M}} = diag\Big\{ 0,0, \frac{MN-1}{2 \pi^2 K^2} \Big\}
\eeq
For $M,N >1$ the third eigenvalue is positive, so the ABJ/ABJM theory is an attractor along the $y_2$--direction. 

Solving the degeneracy of null eigenvalues requires computing the matrix of second derivatives. In particular,
looking at the $y_1$--direction we find 
\beq
\frac{\pa^2 \b_{y_1}}{\pa y_1^2} = \frac{1-M N }{2 \pi^2 K}
\eeq
Since it is non--vanishing, the $y_1$ coordinate is a line of instability. Therefore, when perturbed by a 
$SU(2)_A \times SU(2)_B$ violating operator the system leaves the ABJM fixed point and flows to a 
less symmetric fixed point along a RG trajectory. 

\vskip 10pt
We now generalize the analysis to the case of different CS levels. In this case we refer to the surface
of fixed points in Fig. 5 where for clearness only half of the ellipsoid has been drawn.  
The black line corresponds to ${\cal N}=2$ superconformal theories with $h_1 = h_2 =0$, where the green 
point is the $SU(2)_A \times SU(2)_B$ invariant model. The red point is instead the ${\cal N}=3$ 
superconformal theory.  

From eq. (\ref{gamma}) we see that in the $h_1=h_2$ subsector we have $\g_{A^1} = \g_{A^2}$. 
As a consequence, all the beta--functions are equal and the RG flow equations simplify to
\beq
\frac{d h_i}{d h_j} = \frac{h_i}{h_j}
\eeq
In the three dimensional parameter space $(h_1 = h_2 , h_3, h_4)$,
solutions are all the straight lines passing through the origin and intersecting the ellipsoid.

\begin{figure}
  \center
\label{fig:ellipsoidflow}
\includegraphics[scale=0.5, angle=-90]{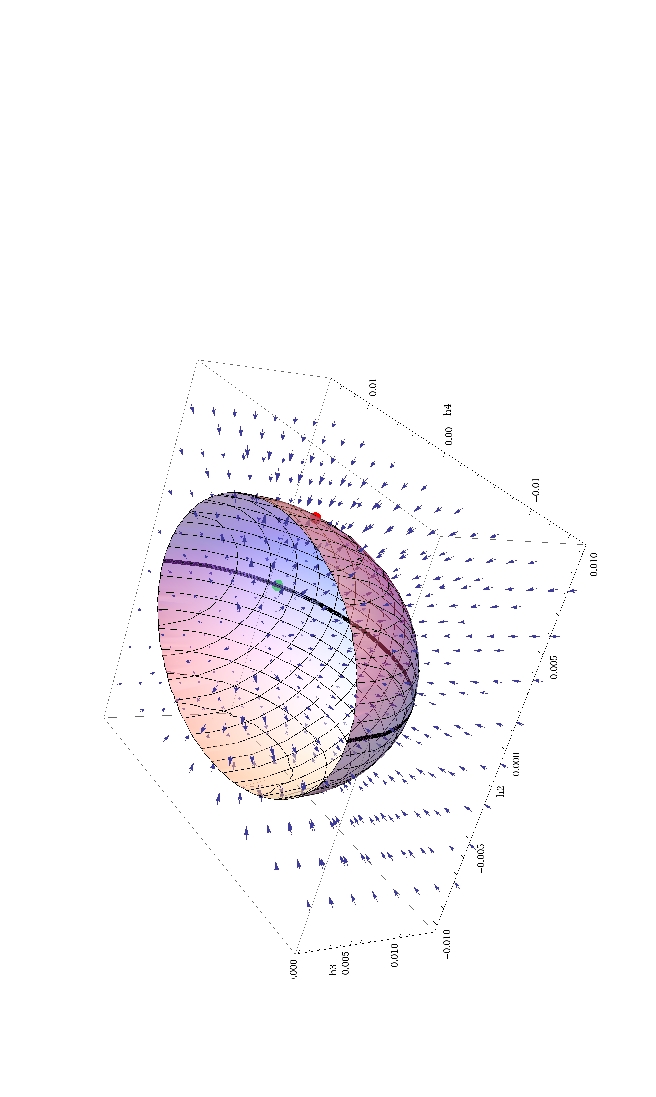}
\caption{The ellipsoid of fixed points and the RG flows for ${\cal N}=2$ theories in the space of
  couplings  $(h_1=h_2, h_3, h_4)$.  Arrows point towards IR directions. 
  The parameters are $K_1 = 150, \, K_2 = 237, \, M= 30$ and $N=43$. }
\end{figure} 

Infrared flows can be easily studied by plotting the vector $(-\b_{h_1}, -\b_{h_3}, -\b_{h_4})$ in each 
point. The result is given in Fig. 5 where it is clear that the entire surface is globally IR stable. 

In order to study the local behavior of the system in proximity of a given fixed point, we compute the 
stability matrix at the point (\ref{surface}) and diagonalize it. Surprisingly, the 
eigenvalues turn out to be independent of the particular point on the surface 
\bea
{\cal M} = diag\left\{0, 0,\frac{K_1^2+4 K_1 K_2+K_2^2+2 \left(K_1^2+K_1K_2+K_2^2\right) M N}{4 K_1^2 K_2^2 \pi ^2}\right\}
\eea
The two null eigenvalues characterize directions of instability. In fact, we can solve the degeneracy
by computing the matrix of second derivatives respect to the corresponding eigenvectors. 
It turns out that in all cases the beta functions have a parabolic behavior along those directions and the
system is unstable.

For example, at the ${\cal N}=2$, $SU(2)_A\times SU(2)_B$ invariant fixed point (green dot in the Figure), 
these eigenvectors are $\{0,1,1\}$ and $\{1,0,0\}$, which are precisely the directions 
$h_3=h_4$ and $h_1$, tangent to the surface at that point. It is clear from Fig. 5 that if we perturb the system 
along these directions it will intercept a RG trajectory which leads it to another fixed point. 

\vskip 10pt

The stability properties of the $\b$--deformed theories are easily inferred from the previous discussion. In fact, 
performing the following rotation of the couplings
\bea
h \cos( \p \b) = \frac{x}{2},\qquad
h \sin( \p \b) = \frac{y}{2}
\eea
the condition (\ref{deformations}) for vanishing beta--functions  becomes
\bea
\frac14 (M N - 1) x^2 +  \frac14 (M N + 1) y^2  =  
\frac12 (2MN+1)\left( \frac{1}{K_1^2} + \frac{1}{K_2^2} \right) + \frac{MN +2}{K_1 K_2} 
\ena
This is exactly the ellipse (\ref{nullbeta2}) of the undeformed case. Therefore,  the infrared stability properties of this curve 
are precisely the ones discussed before.

\subsection{Theories with flavors}

We now turn to flavored theories introduced in Section 4.2.
The form of the stability matrix is quite cumbersome, but we can analyze
the effects of the interactions with flavor multiplets by studying particular
examples.

As the simplest case we consider the class of theories described by the
superpotential (\ref{eqn:superpot}}) where only $\l_i$ couplings have been turned on. 
The $\b$-functions of the theory split into two completely
decoupled sectors: The former is the four dimensional space of
couplings $h_1$, $h_2$, $h_3$, $h_4$, whose stability was addressed in
the previous subsection; the latter is the three dimensional space of
$\l_i$ couplings.

Looking at the $\l_i$ sector, nontrivial solutions to $\b_i=0$  
describe a curve of fixed points given by expressing $\l_1$ and $\l_2$
as functions of $\l_3$ (see eqs. (\ref{eqn:finiteness})). It is the two--branch
curve of Fig. 6. 
The most general solution includes also isolated points where either
$\l_1$ or $\l_2$ vanish. 

Drawing the vector $(-\b_{\l_1}, -\b_{\l_2},-\b_{\l_3})$
in each point of the parameter space we obtain the RG flow configurations as given
in Fig. 6. It is then easy to see that the isolated fixed points are always unstable 
since the RG flows drive the theory to one of the two
branches in the IR. 

This behavior can be also inferred from the structure of the stability matrix. In fact,
one can check that when evaluated on the curve the matrix has two
positive eigenvalues, whereas when evaluated at the isolated solutions it has 
negative eigenvalues.
\begin{figure}
  \center
  \includegraphics[scale=0.5]{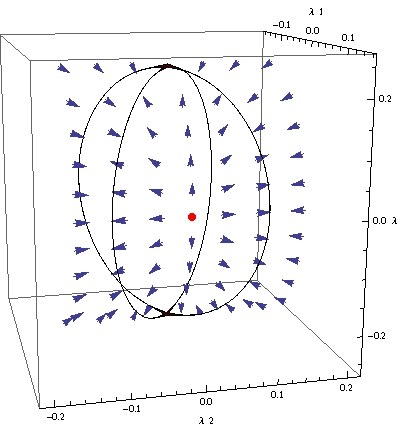}
  \caption{A sketch of the RG flow for the $\l_i$ couplings only. A
  curve of fixed points is shown, which is IR stable. The red dot
  represents an isolated IR unstable fixed point. Here, the parameters are $K_1 =  K_2 = 20, \, M= N=10$, $N_f = N^\prime_f = 1$. }
  \label{fig:lambda}
\end{figure}
As before, theories living on the curve have directions of local
instability  signaled by the presence of a null eigenvalue  
which can be solved at second order in the derivatives. The direction of instability 
is tangent to the curve.

Finally, we consider the more complicated case of theories with 
superpotential (\ref{eqn:superpot}) where only the $\a_i$ couplings are 
non--vanishing. This time the $\b$--functions for the $h_i$ sector do not 
decouple from the $\b$--functions of the $\a_i$ sector and the analysis
of fixed points becomes quite complicated. 

In order to effort  the calculation we restrict to the class of $U(N) \times U(N)$ theories 
(therefore $N_f = N^\prime_f$) with $|K_1| = |K_2|$. This allows to choose 
$\a_i$ all equal to $\a$. Moreover, we set $h_1 = h_2$ and $y_1=0$ in (\ref{y}). 
The spectrum of fixed points and the RG trajectories are then studied in the three-dimensional 
space of parameters $(\a, h_1, y_2)$.

The $\b$--functions vanish for vanishing couplings (free theory) and for $\g_{A^1} = \g_{Q_1} = 0$. 
Nontrivial solutions for $\a$ are obtained from $\g_{Q_1} = 0$. Using eqs. (\ref{gamma}), for real couplings
we find
\beq
\a = \pm \sqrt{\frac{2N^2 + NN_f +1}{2N^2 K_1^2}}
\label{alpha}
\eeq
Fixing $\a$ to be one of the three critical values (zero or one of these two values) we can solve 
$\g_{A^1} = 0$. As in the previous cases this describes an ellipse on the $(h_1,y_2)$ plane localized 
at $\a= {\rm const}$. For theories with $K_1 = K_2$ the configuration of fixed points is given in Fig. 7 
where we have chosen to draw only half ellipses.

\begin{figure}
  \center
\label{fig:alpha-1}
\includegraphics[scale=0.3]{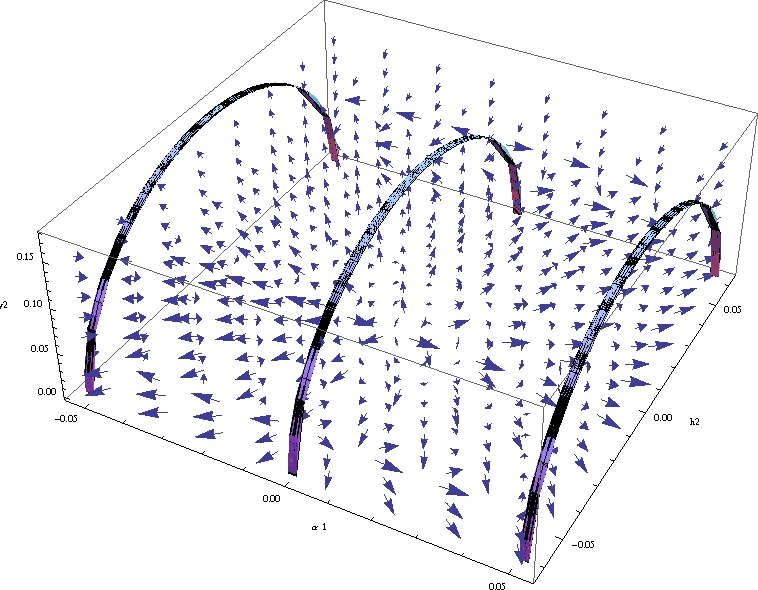}
\caption{The three ellipses of fixed points and the RG flows for ${\cal N}=2$ theories with $\a$ couplings turned on. 
Arrows point towards IR directions.  The parameters are $K_1 =  K_2 = 20, \, M= N=10$, $N_f = N^\prime_f = 1$. }
\end{figure} 

Renormalization group flows are obtained by plotting the vector $(-\b_{\a}, -\b_{h_1},- \b_{y_2})$.
The stability of fixed points is better understood by projecting RG trajectories on orthogonal planes. Looking for
instance at the $h_1=0$ plane we obtain the configurations in Fig. 8 where the red dots indicate the origin and the 
intersections of the three ellipses with the plane. 

\begin{figure}
  \center
\label{fig:alpha-2}
\includegraphics[scale=0.7]{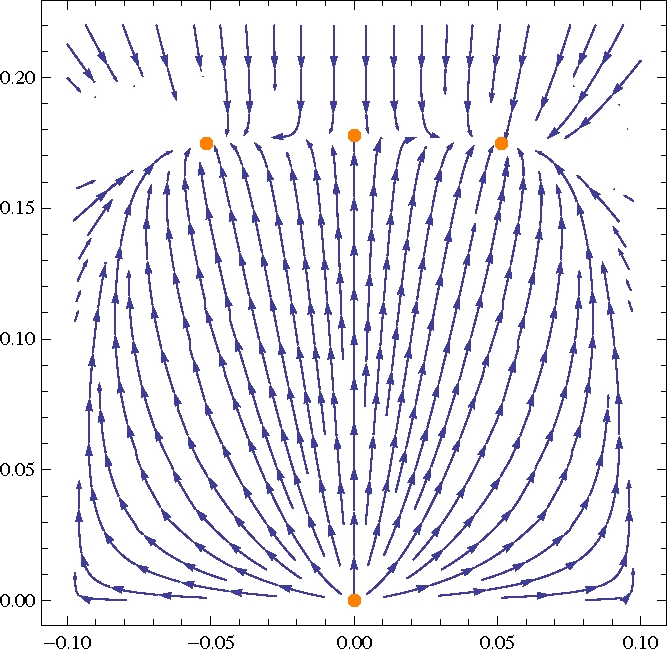}
\caption{RG trajectories on the $h_1=0$ plane for ${\cal N}=2$ theories with $\a$ couplings turned on. 
Arrows point towards IR directions.  The parameters are $K_1 =  K_2 = 20, \, M= N=10$, $N_f = N^\prime_f = 1$.}
\end{figure} 

From this picture we immediately infer that the free theory is an IR unstable fixed point since the system is 
always driven towards nontrivial fixed points. Among them, the ones corresponding to $\a \neq 0$ are
attractors, whereas $\a=0$ does not seem to be a preferable point for the theory. In fact, it is reached 
flowing along the $\a=0$ trajectory, but as soon as we perturb the system with a marginal operator 
corresponding to $\a \neq 0$ it will flow to one of the two nontrivial points. 
We conclude that if we add flavor degrees of freedom the system requires a nontrivial interaction with
bifundamental matter in order to reach a stable superconformal configuration in the infrared region.

\section{Conclusions} 

In this paper we have investigated the spectrum of superconformal fixed points of a large class 
of ${\cal N}=2$, $U(N) \times U(M)$ Chern--Simons theories with bifundamental matter, flavored and not 
flavored.   

We have quantized the theories in a manifest ${\cal N}=2$ superspace setup and evaluated the 
beta--functions perturbatively, at the first nontrivial order. 
We have then determined the whole spectrum of fixed points by setting the betas to zero,
studied the RG flows and the stability properties of the solutions. Choosing the CS levels sufficiently 
large compared to the rank of the gauge groups all the solutions fall inside the perturbative regime
and we are allowed to investigate the IR dynamics around fixed points perturbatively. 

In all cases we have found compact surfaces of fixed points.
They contain as non--isolated solutions, points corresponding to ${\cal N}=6$ ABJ/ABJM theories 
with and without flavors and theories corresponding to turning on a Romans mass in the dual 
supergravity description. 

In a neighborhood of the surface the RG trajectories are straight lines intersecting it. 
Infrared RG flows always point towards the surface which is then globally stable.  
However, a local instability is present and can be understood as follows.
Around a given fixed point the addition of a non--exactly marginal operator drives the system out of the
fixed point. If it happens along the RG trajectory which intersects the surface at that point the system
will flow back to the original superconformal point. But if this does not happen, the perturbed system
meets another RG trajectory which leads it to an infinitesimally closed but different fixed point on
the surface. 

In particular, the ABJM fixed point is stable only respect to $SU(2) \times SU(2)$ invariant perturbations. 
Any other perturbation drives the system to a less symmetric superconformal point.   

When interacting flavors are present, it comes out that IR stability is favored by a 
non--trivial interaction with the bifundamental matter other than with the gauge fields. 

Our analysis could be extended to different classes of theories for which a perturbative investigation of
the IR region makes sense. One example is the class of ${\cal N}=1$ theories introduced in \cite{GT} and 
corresponding to splitting the two CS levels in the ABJ/ABJM action written in ${\cal N}=1$ superspace
formalism.

\vskip 25pt
\section*{Acknowledgements}
\noindent 

Marco Bianchi thanks the Galileo Galilei Institute for hospitality during the completion of this
paper. This work has been supported in part by INFN and PRIN prot.20075ATT78-002.

\vfill
\newpage
\appendix
\section{Notations and conventions}
\label{sec:notations}

In this Appendix we list a number of conventions which are useful for
understanding the technical part of the paper.
   
The world-volume metric is $g^{\mu\nu} = diag(-1,+1,+1)$ with index
range $\mu=0,1,2$. We use Dirac matrices $(\gamma^\mu)_\alpha{}^\beta
= (i \sigma^2, \sigma^1, \sigma^3)$ satisfying $\gamma^\mu \gamma^\nu
= g^{\mu\nu} + \epsilon^{\mu\nu\rho} \gamma_\rho$.

The fermionic coordinates of $\mathN=2$ superspace are two real
two-component spinors $\theta_i, i=1,2$ which we combine into a
complex two spinor
\bea
  \th^\a = \frac{1}{\sqrt{2}} ( \th_1^\a + i \th_2^\a ) \qquad
  , \qquad \thb^\a = \frac{1}{\sqrt{2}} ( \th_1^\a - i \th_2^\a )
\eea

Indices are raised and lowered according to $\theta^\alpha = C^{\a\b} \th_\b$, 
$\th_\a = \th^\b C_{\b\a}$, with $C^{12} = -C_{12} = i$.  
We have
\beq
  \th_\a \th_\b = C_{\b\a} \th^2 \qquad , \qquad
  \th^\a \th^\b = C^{\b\a} \th^2
\eeq
and likewise for $\bar{\theta}$ and derivatives.

Supercovariant derivatives and susy generators are
\bea
  D_\a = \pa_\a + \frac{i}{2} \thb^\b \pa_{\a\b} = \frac{1}{\sqrt{2}}
  ( D_\a^1 -i D_\a^2 ) \qquad &,& \qquad \Db_\a = \bar\pa_\a
  + \frac{i}{2} \th^\b \pa_{\a\b} = \frac{1}{\sqrt{2}} ( D_\a^1 +i
  D_\a^2 ) \non \\ Q_\a = i ( \pa_\a - \frac{i}{2} \thb^\b \pa_{\a\b}
  ) \qquad &,& \qquad \bar Q_\a = i (\bar\pa_\a
  - \frac{i}{2} \th^\b \pa_{\a\b})
\eea
with the only non-trivial anti-commutators
\beq
  \{D_\a , \Db_\b\} = i \pa_{\a\b} \qquad , \qquad \{Q_\a , \bar
  Q_\b \} = i \pa_{\a\b}
\eeq
We use the following conventions for integration
\beq
  d^2\th \equiv \frac{1}{2} d\th^\a d\th_\a \qquad , \qquad
  d^2\thb \equiv \frac{1}{2} d\thb^\a d\thb_\a \qquad , \qquad
  d^4\th \equiv d^2\th \, d^2\thb
\eeq
such that
\beq
  \int d^2\th \, \th^2 = -1 \qquad
  \int d^2\thb \, \thb^2 = -1 \qquad
  \int d^4\th \, \th^2 \thb^2 = 1
\eeq
 
The components of a chiral and an anti-chiral superfield,
$Z(x_L,\theta)$ and $\bar{Z}(x_R,\bar{\theta})$, are a complex boson
$\phi$, a complex two-component fermion $\psi$ and a complex auxiliary
scalar $F$. Their component expansions are given by
\bea
  Z = \phi(x_L) + \th^\a \psi_\a(x_L) - \th^2 \,
  F(x_L) \non \\ \bar{Z} = \bar{\phi}(x_R)
  + \thb^\a \bar{\psi}_\a(x_R) - \thb^2 \, \bar{F}(x_R)
\eea
where
\bea
  x_L^\mu = x^\mu + i \theta \gamma^\mu \bar{\theta} \non \\
  x_R^\mu = x^\mu - i \theta \gamma^\mu \bar{\theta}
\eea
The components of the vector superfield $V(x,\theta,\bar{\theta})$ in
Wess-Zumino gauge ($V| = D_\a V| = D^2 V| = 0$) are the gauge field $A_{\a\b}$, a complex
two-component fermion $\lambda_\a$, a real scalar $\sigma$ and an
auxiliary scalar $D$, such that
\begin{equation}
\label{eqn:WZgauge}
  V = i \, \th^\a \thb_\a \, \sigma(x)
    + \th^\a \thb^\b \, A_{\a\b}(x)
    - \th^2 \, \thb^\a \bar{\lambda}_\a(x) 
    - \thb^2 \, \th^\a \lambda_\a(x)
    + \th^2 \, \thb^2 \, D(x)
    \; .
\end{equation}      

\bigskip

For $SU(N)$ we use the $N\times N$ hermitian matrix generators $T^a$ ($a=1,\ldots, N^2-1$) 
and for $U(N)$
\beq
T^A = (T^0, T^a) \quad {\rm with} \quad  T^0 = \frac{1}{\sqrt{N}}
\eeq
The generators are normalized as $\Tr T^A T^B = \delta^{AB}$. 

Completeness implies 
\bea
&U(N):& \qquad \Tr A T^A \Tr B T^A = \Tr A B
\quad , \quad  \Tr A T^A B T^A = \Tr A \Tr B 
\non\\
\non \\
&SU(N):& \qquad \Tr A T^a \Tr B T^a = \Tr A B - \frac{1}{N} \Tr A
\Tr B 
\non \\
&& \qquad  \Tr A T^a B T^a = \Tr A \Tr B - \frac{1}{N} \Tr A B
\eea

\vskip 10pt
Useful integrals for computing Feynman diagrams in momentum space and dimensional regularization
($d= 3 -2\e$) are, at one loop
\bea
  \intk{k} \frac{1}{k^2 (k-p)^2} &=& \frac{1}{8} \frac{1}{|p|} \equiv
  B_0(p) 
  \label{1integral}
  \\ \intk{k} \frac{k_{\a\b}}{k^2 (k-p)^2} &=& \frac12 \, p_{\a\b} \, B_0(p)
\eea
and at two loops
\bea
F(p) \equiv   \intkk{k}{q} \frac{1}{k^2\, q^2\, (p-k-q)^2} = \frac{\G(\e)}{64\pi^2} \sim \frac{1}{64\pi^2}
\, \frac{1}{\e} 
\label{integral}
\eea

\end{document}